\documentclass[11pt,twoside]{article}

\usepackage{edwin}
\usepackage{epsf}
\usepackage{psfig}
\usepackage{lscape}

\begin{document}
\title{Publication Trends in Astronomy: The Lone Author}   %%% Fill in title
\author{Edwin A. Henneken}
\affil{Smithsonian Astrophysical Observatory, 60 Garden Street, Cambridge, MA 02138}

\begin{abstract} %%% Abstract to run on from here.
In this short communication I highlight how the number of collaborators on papers in the main astronomy journals has evolved over time. We see a trend of moving away from single-author papers. This communication is based on data in the holdings of the SAO/NASA Astrophysics Data System (ADS).

The ADS is funded by NASA Grant NNX09AB39G.
\end{abstract}

%\section{Introduction}
This communication illustrates the trend discussed by Mott Greene in the essay ``The demise of the lone author'' (\citet{greene2007}). Trends are likely to be different for different disciplines. As Mott observes: ``In most fields outside mathematics, fewer and fewer people know enough to work and write alone''. In addition to this, in most disciplines large (and often multi-national) collaborations have become more common and even unavoidable, because it is the only way to get sufficient funding.

\begin{figure}[!ht]
  \plotone{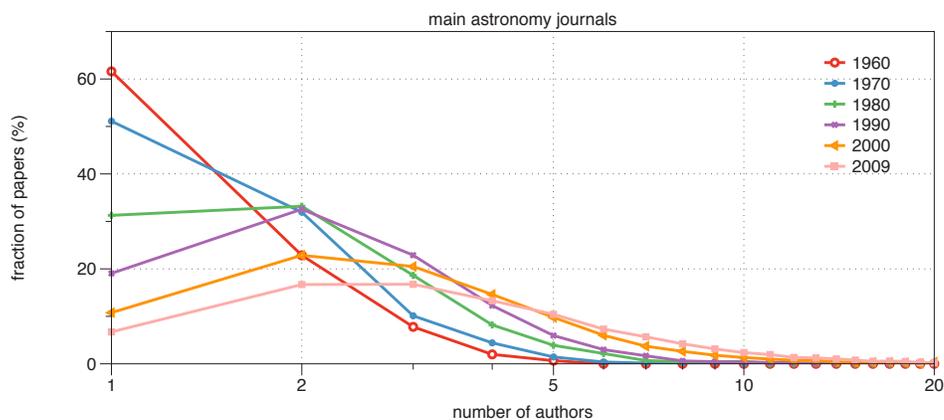}
  \caption{The distribution of the relative frequency of the number of authors per paper in the main astronomy journals for a number of years}
  \label{fig:figure1}
\end{figure}

Figure~\ref{fig:figure1} is an illustration of how the distribution of the number of authors has changed over time in the main astronomy journals ({\it The Astrophysical Journal}, {\it The Astronomical Journal}, {\it Monthly Notices of the R.A.S.} and {\it Astronomy \& Astrophysics}).

\begin{figure}[!ht]
  \plotone{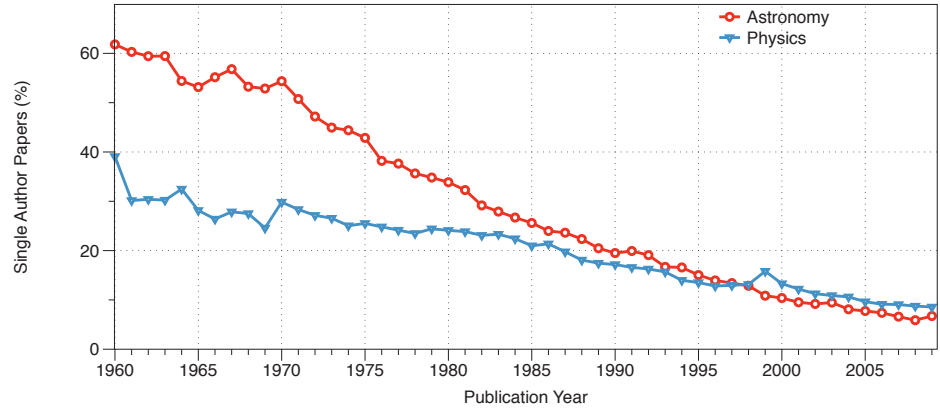}
  \caption{The fraction of papers by single authors in the main astronomy and physics journals}
  \label{fig:figure2}
\end{figure}

Figure~\ref{fig:figure2} highlights the ``demise of the lone author'' by showing the change in the fraction of single author papers in the main astronomy journals. The fraction in the main physics journals ({\it Physical Review}, {\it Nuclear Physics}, {\it Physics Letters}) has been added for comparison.

The drop in the astronomy journals is more dramatic than for the physics journals. A factor of about 10 versus a factor of about 3 or 4.

\end{document}